\documentclass[preprint,tighten,floats,twoside,eqsecnum,aps,epsf,epsfig]{revtex4}
\usepackage{graphicx}

\usepackage{color}
\usepackage{bm}
\newcommand{\GeV}{\makebox{ GeV}}
\newcommand{\beq}{\begin{equation}}
\newcommand{\enq}{\end{equation}}
\newcommand{\beqa}{\begin{eqnarray}}
\newcommand{\beqast}{\begin{eqnarray*}}
\newcommand{\enqa}{\end{eqnarray}}
\newcommand{\enqast}{\end{eqnarray*}}

\def\GeV{\nobreak\,\mbox{GeV}}

\begin{document}

\title{Elastic $\rm{p\bar p}$ Scattering Amplitude at 1.8 TeV and
Determination of Total Cross Section}
 
  \author{A. K. Kohara} 
 \affiliation{Instituto de F\'{\i}sica, Universidade Federal do Rio de
 Janeiro \\
 C.P. 68528, Rio de Janeiro 21945-970, RJ, Brazil   }
\author{E. Ferreira} 
 \affiliation{Instituto de F\'{\i}sica, Universidade Federal do Rio de
 Janeiro \\
 C.P. 68528, Rio de Janeiro 21945-970, RJ, Brazil   }
\author{T. Kodama} 

 \affiliation{Instituto de F\'{\i}sica, Universidade Federal do Rio de
 Janeiro \\
 C.P. 68528, Rio de Janeiro 21945-970, RJ, Brazil   }


\begin{abstract}
~~~~The data on p$\mathrm{\bar p}$ elastic scattering at 1.8 and 1.96 TeV
are analysed in terms of real and imaginary amplitudes, in a treatment with
high accuracy, covering the whole t-range and satisfying the
expectation of dispersion relation for amplitudes and for slopes. A method
is introduced for determination of the total cross section and the other
forward scattering parameters and to check compatibility of E-710, CDF
and the recent D0 data. Slopes $B_R$ and $%
B_I$ of the real and imaginary amplitudes, treated as independent
quantities, influence the amplitudes in the whole t-range and are important
for the determination of the total cross section. The amplitudes
are fully constructed, and a prediction is made of a marked dip in $%
d\sigma/dt$ in the $|t|$ range 3 - 5 GeV$^2$ due to the universal
contribution of the process of three gluon exchange.

\end{abstract}

\maketitle
\section{Introduction \label{sec-introduction}}

The precise knowledge of total cross section and scattering amplitudes in pp
and p\={p} elastic scattering at high energies is essential for understanding
the QCD interactions and hadronic structure, and also for the parametrization
and extrapolation of the total cross section that may pass through the LHC
measurements \cite{TOTEM} and go up to the study of ultra-high energy
phenomena in cosmic rays rays \cite{Auger}. However, at high energies, the
smallness of the ratio $\rho$ of the real to the imaginary parts of amplitudes
at $t=0,$ together with the absence of data for small $|t|$, turn the
extrapolations towards the limit $|t|\rightarrow0$ very delicate. It is of
fundamental importance to characterize well the scattering amplitudes that are
used to determine forward slopes and total cross section.

It is universally understood that the real and imaginary amplitudes in pp and
p$\mathrm{{\bar{p}}}$ elastic scattering at high energies reflect the
non-perturbative QCD dynamics, determined by overall features of the proton
and antiproton structures. Regge-like behavior characterizes the $s$ and $t$
dependences at large $s$ and small $|t|$. There appears a dip or an inflection
point in the differential cross section $d\sigma/dt$, near the occurrence of a
zero in the imaginary part, and the detailed shape around this region is
influenced by the magnitude, sign and form of the real part. An analysis of
the interplay of real and imaginary amplitudes is necessary to reproduce with
accuracy the behavior of the $|t|$ dependence.

An analytical representation for the amplitudes valid for the whole $|t|$
range must contain implicitly the exponential decrease of the amplitudes in
the very forward region, account for their curvatures, zeros, signs and
magnitudes, and also should contain the ingredients that describe the
universal power behavior at large $|t|$ due to the three-gluon exchange
contribution. The determination of the detailed properties of the real and
imaginary parts is crucial for the accurate description of the observed
differential cross section. The analytical forms of the amplitudes used in the
present work extend previous studies at the ISR energies \cite{ferreira1}
based on the Stochastic Vacuum Model \cite{dosch}. More recently, with
additional controls offered by dispersion relations for amplitudes and for
slopes \cite{ferreira2}, the method has been applied to the recent 7 TeV data
from LHC \cite{KEK_2013}.

The present work extends the previous studies to give a high precision
description of the data on p$\mathrm{{\bar{p}}}$ elastic scattering at 1.8 TeV
\cite{Amos,Abe} and 1.96 TeV \cite{ppbar1960}, consistently covering the
forward and the backward regions. This work is particularly opportune in view
of the publication of the new measurements at 1.96 TeV covering a large $|t|$
range by the D0 Fermilab experiment \cite{ppbar1960}. Our framework offers the
opportunity of an investigation of the connection and compatibility of the
previous 1.8 TeV and the recent 1.96 TeV data. It should be stressed here that
in the analysis of the data, we use information from forward dispersion
relations to control parameters of the full description and particularly
emphasize the importance of the difference of the slope parameters, $B_{R}$
and $B_{I}$, of the real and imaginary parts. It is commonly assumed in the
analysis of data that these slopes are the same, but this is wrong
theoretically, fact that is often overlooked due to the smallness of the
$\rho$ parameter. However, to describe consistently the scattering amplitude
for the full $\left\vert t\right\vert $ range, the difference of slopes is
crucial, since a description that covers the large $t $ region constrains the
quantities of the forward range. This is particularly true and important for
the real amplitude that is small with respect to the imaginary part in the
forward direction, but not at large $|t|$. Actually, in our analysis the
differential cross section at high $|t|$ is dominated by the real part
\cite{ferreira1, KEK_2013}.

In the very large $|t|$ domain, the perturbative QCD effects become dominant,
forming a power decreasing tail in the differential cross section, that was
first measured at 27.4 GeV \cite{Faissler}. It is known that this tail is
energy independent and formed by a real contribution due to three-gluon
exchange \cite{DL_3g}, with opposite signs for pp and p$\mathrm{{\bar{p}}}$
(positive in pp and negative in p$\mathrm{{\bar{p}}}$) scattering. In our
approach the universality of the perturbative 3-gluon exchange process is
incorporated explicitly, determining the asymptotic behavior of the real
amplitude. It is thus interesting to investigate the connection of the
measured points at 1.8 and 1.96 TeV with the assumption of the universal tail.
We show that the perturbative amplitude leads to a striking prediction for the
behavior of the cross section. In p$\mathrm{{\bar{p}}}$ scattering, when added
to the non-perturbative positive real part, the perturbative term creates a
third zero, located in the region about $|t| ~ \approx~ 3~ - ~4
~\nobreak\,\mbox{GeV}^{2}$. As the imaginary part is less important in this
domain, a marked dip is caused by this cancellation.

Our treatment of the whole data with overall high precision leads to definite
prediction for the forward scattering parameters in the context of one
analytical form. We thus have a determination of the total cross section
$\sigma$ and of the quantities $\rho$, $B_{I}$ and $B_{R}$, while still
allowing curvatures of the scattering amplitudes. When data are not available
at very low $|t|$, the existence of curvatures prevents accurate analysis in
terms of pure exponential forms, so that the determination of $\sigma$ becomes
model dependent. Based on our experience with other energies, we here avocate
that the determination of the scattering parameters based on the full data is
more reliable, since it incorporates realistic properties of the amplitudes.

We organize the present work as follows. In Sec. 2
we present the analytical forms of the scattering amplitudes that describe the
whole $\left\vert t\right\vert $ range , and the necessary quantities are
defined, with a discussion of the role of the universal perturbative
contribution for large $\left\vert t\right\vert $. In Sec. 3
the analysis of the 1.8-1.96 TeV data with determination of all quantities is
presented. In Sec. 4
we present the amplitudes and compare our results with other theoretical
models. In Sec. 5
we discuss our predictions and proposals.


\section{ General form of full $t$ scattering amplitude\label{general}}

In the treatment of elastic pp and p$\mathrm{\bar{p}}$ scattering in the
forward direction, with amplitudes approximated by pure exponential forms, the
differential cross section is written
\begin{eqnarray}
\frac{d\sigma}{dt}  & =\pi\left(  \hbar c\right)  ^{2}~~\Big\{\Big[\frac
{\rho\sigma}{4\pi\left(  \hbar c\right)  ^{2}}~\mathrm{{e}^{B_{R}~t/2}%
+F^{C}(t)\cos{(\alpha\Phi)}\Big]^{2}}\nonumber\\
& +\Big[\frac{\sigma}{4\pi\left(  \hbar c\right)  ^{2}}~\mathrm{{e}%
^{B_{I}~t/2}+F^{C}(t)\sin{(\alpha\Phi)}\Big]^{2}\Big\}~,}\label{diffcross_eq}%
\end{eqnarray}
where $t\equiv-|t|$ and we assume different values for the slopes $B_{I}$ and
$B_{R}$ of the imaginary and real amplitudes. In the following discussion, we
use the unit system where $\sigma$ is in mb(milibarns) and energy in GeV, so
that $\left(  \hbar c\right)  ^{2}= 0.389$ mb GeV$^{2}$.

The Coulomb amplitude $F^{C}(s,t)$ enters for pp$/$p$\mathrm{\bar{p}}$ with
the form
\begin{equation}
F^{C}(s,t)e^{i\alpha~\Phi(s,t)}=(-/+)~\frac{2\alpha}{|t|}e^{i\alpha~\Phi
(s,t)}~F_{\mathrm{proton}}^{2}(t)~,\label{coulomb}%
\end{equation}
where $\alpha~$is the fine-structure constant, $\Phi(s,t)$ is the Coulomb
phase and the proton form factor is written
\begin{equation}
F_{\mathrm{proton}}(t)=[0.71/(0.71+|t|)]^{2}~.\label{ff_proton}%
\end{equation}
Contradicting expectations from dispersion relations \cite{ferreira2}, in
usual treatments of the data no distinction is made between $B_{R}$ and
$B_{I}$ slopes, and $B_{R}\neq B_{I}$ requires a more general expression for
the Coulomb phase \cite{KEK_2013}, which is used in the present work.

In elastic pp and p$\mathrm{\bar p}$ scattering at all energies above $\sqrt
s=$ 19 GeV, the real and imaginary amplitudes have zeros located in ranges
$|t|\approx(0.1-0.3)$ GeV$^{2}$ and $|t|=(0.5-1.5)$ GeV$^{2}$ respectively,
and the use of exponential forms beyond a limited forward range leads to
inaccurate determination of the characteristic forward scattering parameters
$\sigma$, $\rho$ , $B_{I}$ and $B_{R}$. To obtain precise description of the
elastic $d\sigma/dt$ data for all $|t|$, we introduce amplitudes with forms
\cite{ferreira1, KEK_2013}
\begin{equation}
T_{R}(s,t)=\alpha_{R}(s)\exp(-\beta_{R}(s)|t|)+\lambda_{R}(s)\Psi_{R}%
(\gamma_{R}(s),t)+\sqrt{\pi}F^{C}(t)\cos(\alpha\Phi)~,\label{real}%
\end{equation}
and
\begin{equation}
T_{I}(s,t)=\alpha_{I}(s)\exp(-\beta_{I}(s)|t|)+\lambda_{I}(s)\Psi_{I}%
(\gamma_{I}(s),t)+\sqrt{\pi}F^{C}(t)\sin(\alpha\Phi)~,\label{imag}%
\end{equation}
with the shape functions
\begin{equation}
\Psi_{K}(\gamma_{K}(s),t)=2~\mathrm{e}^{\gamma_{K}}~\bigg[{\frac
{\mathrm{e}^{-\gamma_{K}\sqrt{1+a_{0}|t|}}}{\sqrt{1+a_{0}|t|}}}-\mathrm{e}%
^{\gamma_{K}}~{\frac{e^{-\gamma_{K}\sqrt{4+a_{0}|t|}}}{\sqrt{4+a_{0}|t|}}%
}\bigg]~,\label{cd1_s}%
\end{equation}
where $K=R$ for the real amplitude and $K=I$ for the imaginary amplitude. We
here have eight quantities $\alpha_{I}$, $\beta_{I}$, $\gamma_{I}$,
$\lambda_{I}$, $\alpha_{R}$, $\beta_{R}$, $\gamma_{R}$, $\lambda_{R}$ that
determine the non-perturbative amplitudes. $\gamma_{K}$ is dimensionless,
while $\alpha_{K} $, $\lambda_{K}$ and $\beta_{K}$ are in GeV$^{-2}$. These
forms have been developed in the application of the Stochastic Vacuum Model to
pp and p$\mathrm{\bar{p}}$ elastic scattering \cite{ferreira1}, and the fixed
quantity $a_{0}=1.39$ GeV$^{-2}$ is related to the square of the correlation
length of the gluon vacuum expectation value ($~a~=(0.2-0.3)$ fm) \cite{dosch}.

From the above expression, we can express the total cross section
$\sigma\left(  s\right)  , $the ratio $\rho$ of the real to imaginary
amplitudes, the slopes $B_{R,I}$ of the amplitudes at $t=0,$ and the
differential cross section $d\sigma/dt$ as
\begin{equation}
\sigma(s)=4\sqrt{\pi}\left(  \hbar c\right)  ^{2}~(\alpha_{I}(s)+\lambda
_{I}(s))~,
\end{equation}%
\begin{equation}
\rho(s)=\frac{T_{R}(s,t=0)}{T_{I}(s,t=0)}=\frac{\alpha_{R}(s)+\lambda_{R}%
(s)}{\alpha_{I}(s)+\lambda_{I}(s)}~,
\end{equation}%
\begin{eqnarray}
B_{K}(s) &=\frac{1}{T_{K}(s,t)}\frac{dT_{K}(s,t)}{dt}\Big|_{t=0}~\nonumber\\
         &=\frac{1}{\alpha_{K}(s)+\lambda_{K}(s)}\Big[\alpha_{K}(s)\beta_{K}%
(s)+\frac{1}{8}\lambda_{K}(s)a_{0}\Big(6\gamma_{K}(s)+7\Big)\Big]~,
\end{eqnarray}%
\begin{equation}
\label{dsigdt}\frac{d\sigma}{dt}=\left(  \hbar c\right)  ^{2}~|T_{R}%
(s,t)+iT_{I}(s,t)|^{2}~.
\end{equation}
We have thus defined the form of the amplitudes for all $t$ at each energy.
The parameters must be determined by a phenomenological analysis of the data,
with control from dispersion relations for amplitudes and for slopes. The
forms of Eqs. (\ref{real}), (\ref{imag}), (\ref{cd1_s}) are able to describe
the imaginary and real amplitudes at all energies, with consistency in their
features (magnitudes, signs, locations of zeros), and with smoothness in the
energy dependence of the parameters. Values of $\rho$ and $B_{R}$ must be
related with $\sigma$ and $B_{I}$ respecting dispersion relations. This
description represents the non-perturbative QCD dynamics that is responsible
for the soft elastic hadronic scattering. They effectively account for the
terms of Regge and eikonal phenomenology that determine the process up to
$|t|$ ranges up to about $|t|\approx2.0$ GeV$^{2} $.

This representation of the scattering amplitudes has been used successfully to
reproduce the data at ISR \cite{ferreira1} and LHC energies \cite{KEK_2013}.
In these applications, it was found that the imaginary amplitude presents one
zero located in the range (0.5 - 1.5 GeV$^{2}$), while the real amplitude
presents one zero at low $|t|$ ($|t|<0.3$ GeV$^{2}$), according to a theorem
by Martin \cite{Martin}, and a second zero whose location determines the shape
of $d\sigma/dt$ around the dip (or inflection point). As a general behavior,
we have that the imaginary part, Eq. (\ref{imag}), is negative and the real
part, Eq. (\ref{real}) is positive for $|t|$ larger than $1.5$ GeV$^{2}$.
These simple features are general and all data are described accurately.

It is observed that after the dip (or inflection point) the behavior of the
differential cross sections becomes increasingly energy independent. The
elastic pp experiment at $\sqrt{s}=27$ GeV \cite{Faissler} has measured the
range from 5.5 to 14.2 GeV$^{2}$ and these are the only measurements at such
high values of $|t|$. This distribution at high $|t|$ shows remarkable
universality: at all energies $\sqrt{s}$ = 23.5, 30.7, 44.7, 52.8 and 62.5 GeV
, namely at all energies where measurements have reached the intermediate $|t|$
region, $d\sigma/dt$ approaches the same set of points of the 27.4 GeV experiment.

The observed $d\sigma/dt$ at the tail has a dependence of form $1/|t|^{8}$,
and has been explained by Donnachie and Landshoff \cite{DL_3g} as being of
perturbative origin, due to a contribution of three gluon exchange. This term
is real and has an amplitude of the form
\begin{equation}
A(s,t)_{ggg}=-\frac{N}{|t|}\frac{5}{54}\Big[4\pi\alpha_{s}(|\bar{t}|)\frac
{1}{m^{2}(|\bar{t}|)+|\bar{t}|}\Big]^{3}~,
\end{equation}
where
\begin{equation}
\alpha_{s}(|\bar{t}|)=\frac{4\pi}{(11-\frac{2}{3}N_{f})\Big[\log
\Big(\frac{m^{2}(|\bar{t}|)+\bar{t}}{\Lambda^{2}}\Big)\Big]}%
\end{equation}
is the strong coupling constant and $m(|t|)$ is the gluon effective mass
\cite{Cornwall}. The factor 3 in $~\bar{t}\approx(\sqrt{t}/3)^{2}$ comes from
the assumption that each gluon carries one third of the momentum. The
normalization factor $N$ is negative and determined by the nucleon structure.
To extend our description to include the very high $|t|$ range of this form,
we introduce a term $R_{ggg}(t)$ , writing
\begin{equation}
T_{R(\mathrm{{tail})}}(s,t)=\alpha_{R}(s)\exp(-\beta_{R}(s)|t|)+\lambda
_{R}(s)\Psi_{R}(\gamma_{R}(s),t)+\sqrt{\pi}F^{C}(t)\cos(\alpha\Phi
)+R_{ggg}(t)~,
\end{equation}
where $R_{ggg}(t)$ is chosen so that the differential cross section to be
dominated by a term of the form $|t|^{-8}$ for large $|t|$ values (say above
2.5 GeV$^{2}$), while for small $|t|$ the amplitude stays determined by the
original non-pertubative expression. The perturbative three-gluon exchange has
opposite signs for pp and p$\mathrm{\bar{p}}$ scattering, being positive for
pp and negative for p$\mathrm{\bar{p}}$. We then take the following
expression,%
\begin{equation}
R_{ggg}(t)~\equiv~\pm~0.45~t^{-4}(1-e^{-0.005|t|^{4}})(1-e^{-0.1|t|^{2}%
})~,\label{tail}%
\end{equation}
where the signs $\pm$ apply to the pp and p$\mathrm{\bar{p}}$ amplitudes
respectively. The factor $0.45$ is chosen to reproduce the Faissler
measurements and the last two factors are written to suppress smoothly the
perturbative contribution for small $|t|.$ The transition region from 2 to 5
GeV$^{2}$ contains information on the nature and superposition of
non-perturbative and perturbative contributions, and must be investigated,
both experimentally and theoretically. In the ${\rm{p\bar p}}$ case 
the negative sign may lead to a zero in the real amplitude, with 
interesting consequence for the form of $d\sigma/dt$.

The change in sign of this contribution
for pp and p$\mathrm{\bar{p}}$ amplitudes leads to a very interesting
consequence for $\mathrm{p\bar{p}}$ case, which will be discussed in
Sec.(\ref{sec-amplitudes}).

\clearpage

\section{Analysis of elastic $\mathrm{p\bar{p}}$ data at 1.8 $\rm TeV$ 
\label{sec-data-analysis}} 

The available
experimental data on differential cross section of $\mathrm{p\bar{p}}$ elastic
scattering at 1.8 TeV are

\begin{itemize}
\item N = 51 points in the interval $0.00339 \leq|t| \leq0.627 $ (in GeV$^{2}%
$) from the Fermilab E-710 experiment published by N. Amos et al \cite{Amos}
in 1990.

\item N = 26 points in the interval $0.0035 \leq|t| \leq0.285 $ (in GeV$^{2}
$) from the Fermilab CDF experiment published by F. Abe et al \cite{Abe} in 1994.
\end{itemize}

To these data we may now add the results of the experiment at 1.96 TeV

\begin{itemize}
\item N = 17 points in the interval $0.26 \leq|t| \leq1.20 $ (in GeV$^{2}$)
from the Fermilab D0 experiment published by V. M. Abazov et al
\cite{ppbar1960} in 2012.
\end{itemize}

In order to use the last set together with the former two sets, in this paper
we use the reduction factor $(1.8/1.96)^{0.3232}=0.973$ obtained as correction
of energy effect from 1.96 to 1.8 TeV according to Regge
phenomenology \cite{Regge}. As the $|t|$ range
involved is small we neglect the $|t|$ dependence of this factor. In the
following, we refer to these converted data as "1.96 TeV data".

\begin{figure}[ptb]
\label{data-figures} \includegraphics[width=6.5cm]{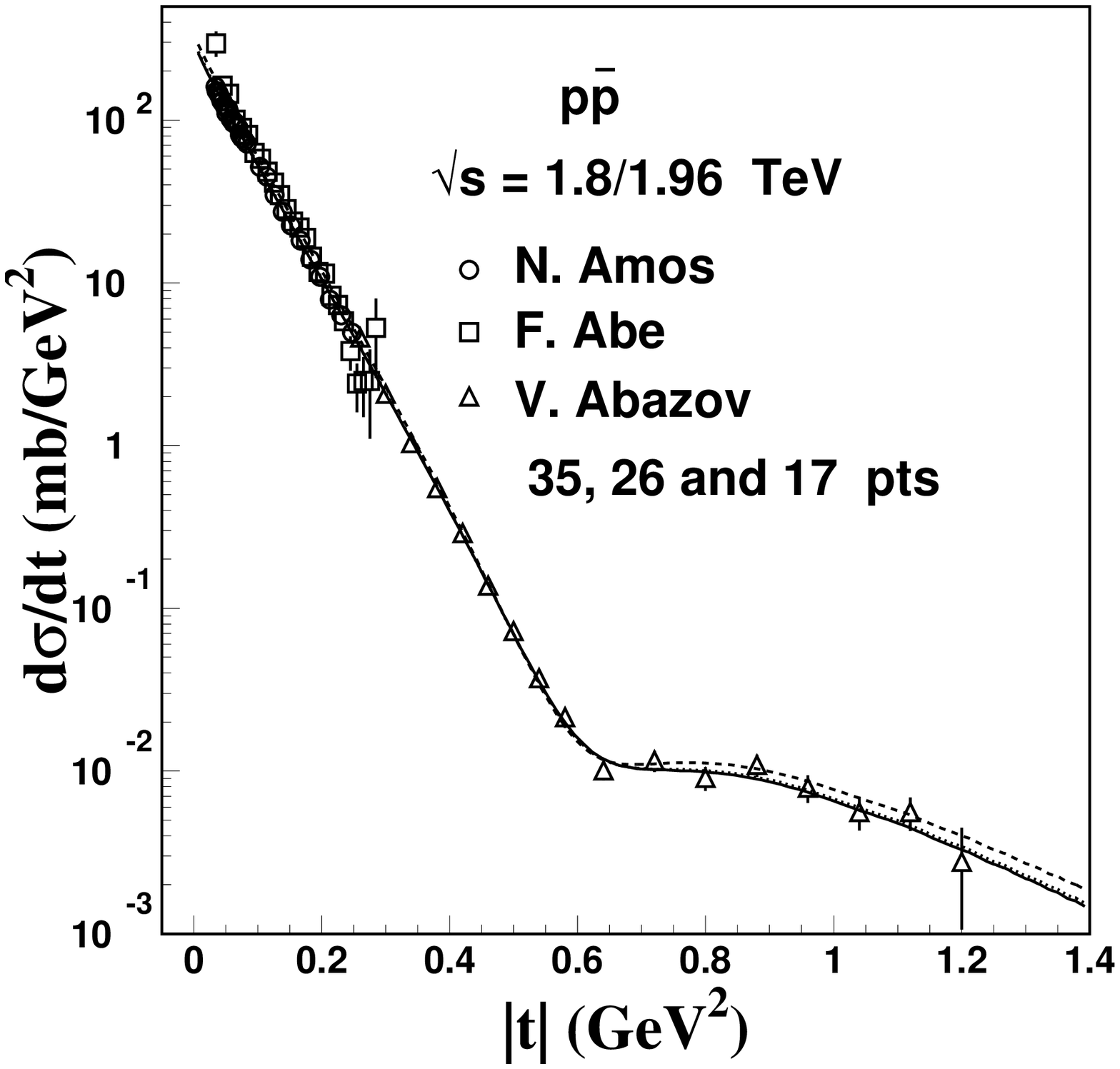}
\includegraphics[width=6.5cm]{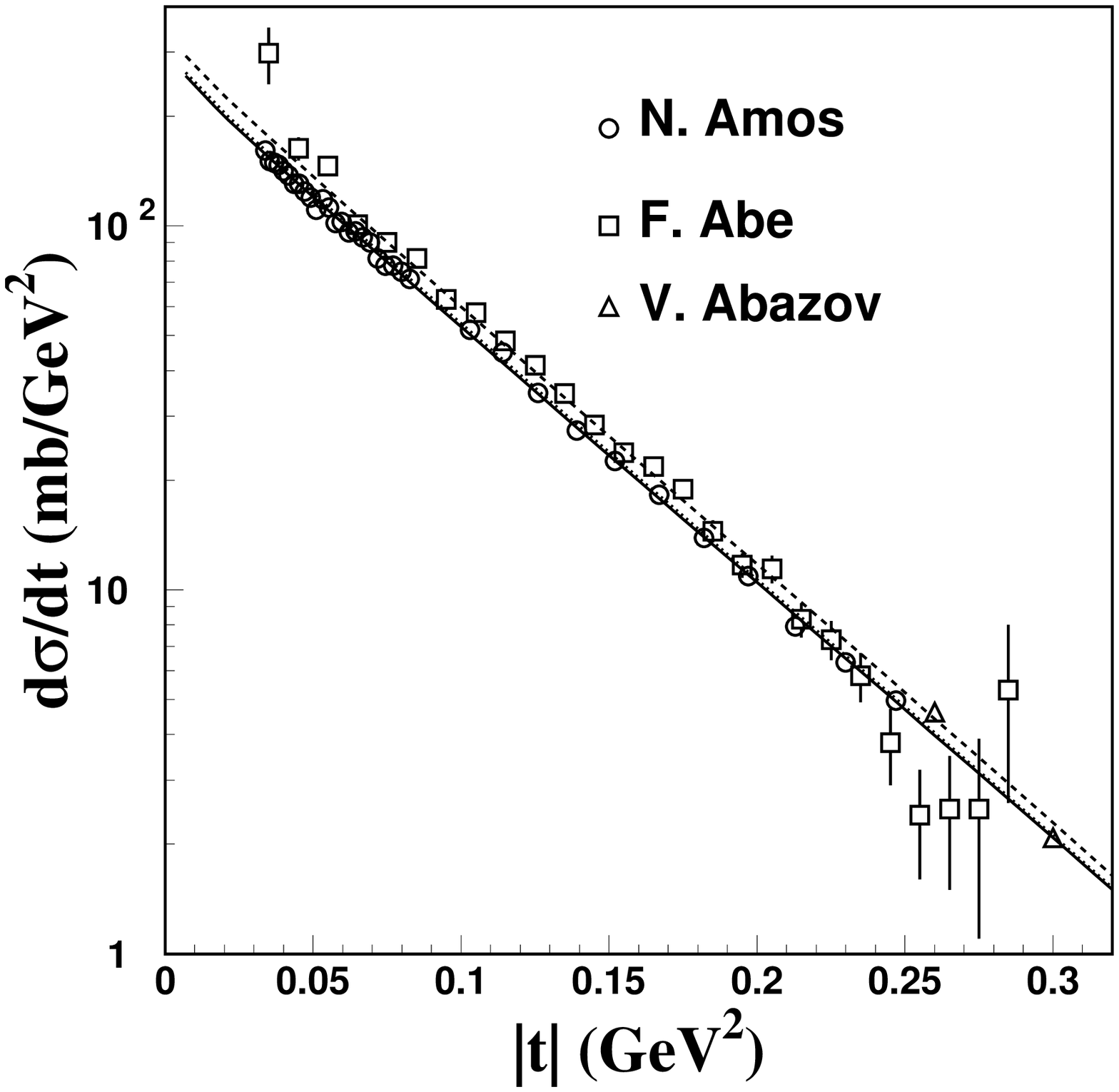}\caption{ Data of p$\bar{\mathrm{p}%
}$ scattering at 1.8 and 1.96 TeV \cite{Amos,Abe,ppbar1960}, taken in the
E-710, CDF and D0 experiments in Fermilab. The D0 data are multiplied by a
reducing factor 0.973 to take into account the energy difference (see the
text). The E-710 data \cite{Amos} are restricted to the first 35 points (open
circles) due to superposition with the recent D0 data (open triangles) of the
same experimental group. The plot in the RHS, concentrated in the forward
part, exhibits clearly the known discrepancy between the two experiments in
the low $|t|$ region. The solid, dashed and dotted lines represent
respectively our best descriptions for datasets I, II and III constructed from
the combination of three data points available, as described in the text. The
dotted line is hidden under the solid line. }%
\end{figure}

The data are shown in Fig. \ref{data-figures}. They do not cover a low enough
$|t|$ range for a precise treatment in terms of exponential forms for the
amplitudes, or, even less, for the differential cross section. Besides, there
is a discrepancy of values in the data from the two independent experiments,
exhibited in Fig. \ref{data-figures}, that has lead to a 20 year old duplicity
of values for the total cross section, which has seriously affected the
efforts for a global description of the energy dependence of the total cross
section.

We recall values of the scattering parameters that are found in original
papers by experimental groups:
\begin{itemize}
\item E-710 experiment \cite{data_exp1}: $\rho=0.140\pm0.069$ , $B=16.99\pm
0.47~ \nobreak\,\mbox{GeV}^{2}$ , $\sigma=72.8\pm3.1$ mb
\item CDF experiment \cite{data_exp2}: $B=16.98\pm0.25~GeV^{2}$ ,
$\sigma=80.03\pm2.24$ mb
\end{itemize}
In the present work we analyse carefully this duplicity using a full-t
analytic description of the real and imaginary amplitudes with help of the new
large $|t|$ data from the 1.96 TeV experiment. As much as possible, we deal
with all experimental information together in a unified analysis. For this
purpose, we group the data in three different sets.

\begin{figure}[ptb]
\label{experimental-data}
\includegraphics[width=7.5cm]{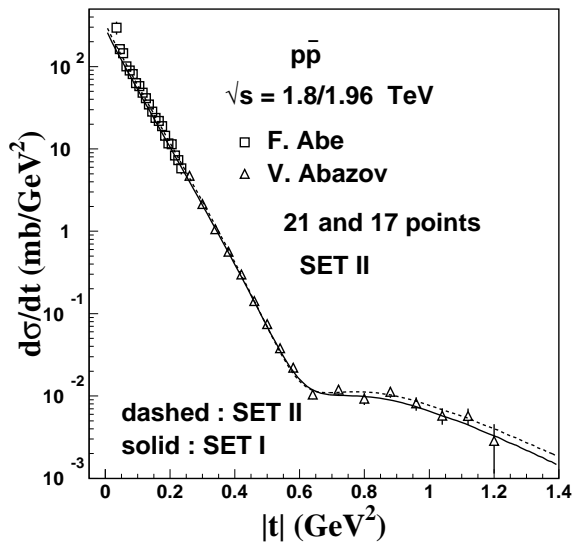}\caption{ HYBRID SET . Combination
of N=21 points from CDF (open squares) with 17 points from D0 (open
triangles). The last 5 points of CDF data (see Fig. \ref{data-figures}) are
excluded, to exhibit more clearly a smooth connection, and this is done also
numerically in fittings with SET II (38 points). The E-710 points do not enter
in this SET II. Solid line: fit of SET I, for comparison; dashed line: fit of
SET II. Although the lines of the two solutions are visually very close, the
limits $|t|\to0$ lead to different values of $\sigma$, given in Table
\ref{fittings} and shown in closeup in Fig. \ref{hybrid-fit}. }%
\end{figure}

\begin{itemize}
\item SET I - The 1.96 TeV data (converted) give a natural and smooth
connection with the E-710 data (basically they come from the same experimental
group); there is some superposition in the extreme ends, where we select the
more recent data, that have smaller error bars. We thus join 35 points with
$3.39 \times10^{-2} \leq|t|\leq0.247$ GeV$^{2}$ from E-710 with 17 points from
D0, to form a combined data SET I (called STANDARD), with N = 52 points in the
range $0.00339 \leq|t| \leq1.2 ~ \nobreak\,\mbox{GeV}^{2}$.

\item SET II - We observe that there is a good convergence of the large $|t|$
end points of the CDF spectrum with the beginning of the recent D0 points.
This is a welcome surprise, and suggests the consistent construction of a
combined set from the two groups, with N = 26 and the N = 17 points,
respectively. Actually, to select points in the range where there is
superposition, and simultaneously to obtain a clearer smooth connection, we
exclude the last 5 CDF points, that present a rather scattered behavior
(observe Fig. 1). We thus build SET II here (called HYBRID), with N = 21 + 17
= 38 points. The assemblage is shown in Fig. \ref{experimental-data}. The
construction of this HYBRID SET II is motivated by the consideration that the
apparently irreconciliable discrepancy between  the E-710 and CDF
experiments that exists in the low and mid $|t|$ range need not imply that
they are incompatible for larger $|t|$. Our description aims at
representations of $d\sigma/dt$ covering all $|t|$ spectrum, and this hybrid
connection is very important.

\item SET III - In a third construction, we investigate what comes out from
our full-$|t|$ description if we put all data together on the same footing,
merging the N = 52 points of SET I with the N = 26 CDF basis. We thus form a
GLOBAL SET III, with N=78 points.
\end{itemize}

We fit $d\sigma/dt$ for the three datasets described above, using Eqs.
(\ref{real}), (\ref{imag}), (\ref{cd1_s}), (\ref{dsigdt}) of our
representation. In the fitting procedure, in principle all 8 parameters are
treated independently to minimize $\chi^{2}$, but we find that some parameters
can be chosen with common values to all datasets without sensitive changes in
the solutions. They are :
\begin{eqnarray}
\alpha_{I}= 11.620\pm0.024 ~\nobreak\,\mbox{GeV}^{-2}, ~\beta_{R}=1.10 ~
\nobreak\,\mbox{GeV}^{-2} ~, ~ \rho=0.141\pm0.002 ~,\nonumber\\
~ B_{I}=16.76\pm0.04 ~ \nobreak\,\mbox{GeV}^{-2}, ~B_{R}=26.24\pm0.39
\nobreak\,\mbox{GeV}^{-2} ~ .\label{basic_parameters}%
\end{eqnarray}

We remark that the usual quantity $B$ (slope of $d\sigma/dt$) is not the same
as $B_{I}$. The relation is
\begin{equation}
B=\frac{B_{I}+\rho^{2} B_{R}}{1+\rho^{2}} ~
\end{equation}
and we then obtain $B=16.94 ~\nobreak\,\mbox{GeV}^{-2}$, remarkably close to
the values of the experimental groups ($16.99\pm0.47$ and $16.98\pm0.25$
$\nobreak\,\mbox{GeV}^{-2}$ for the E-710 \cite{data_exp1} and CDF
\cite{data_exp2} groups respectively).

\begin{figure}[ptb]
\label{hybrid-fit}
\includegraphics[width=6.5cm]{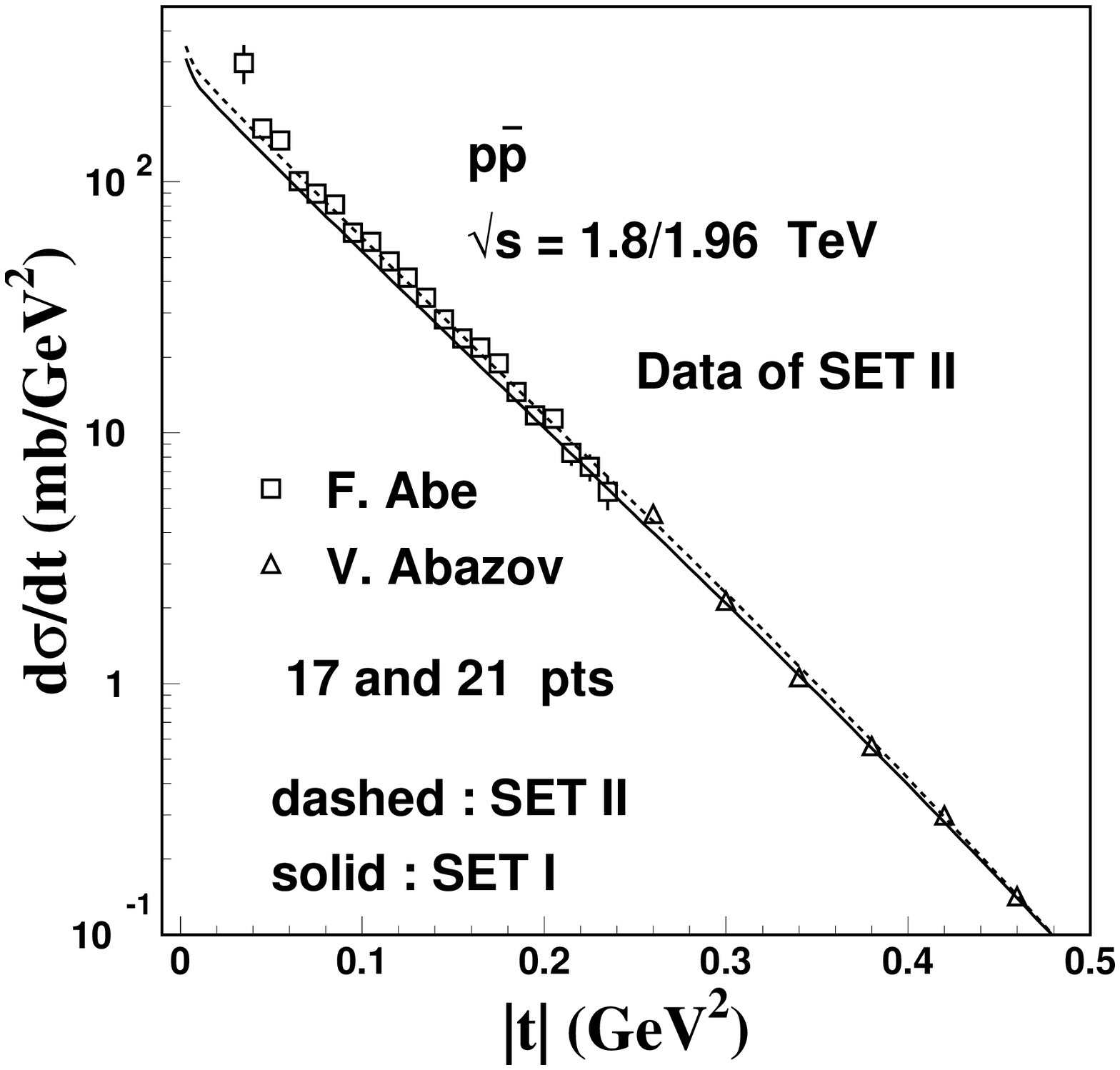}
\includegraphics[width=6.5cm]{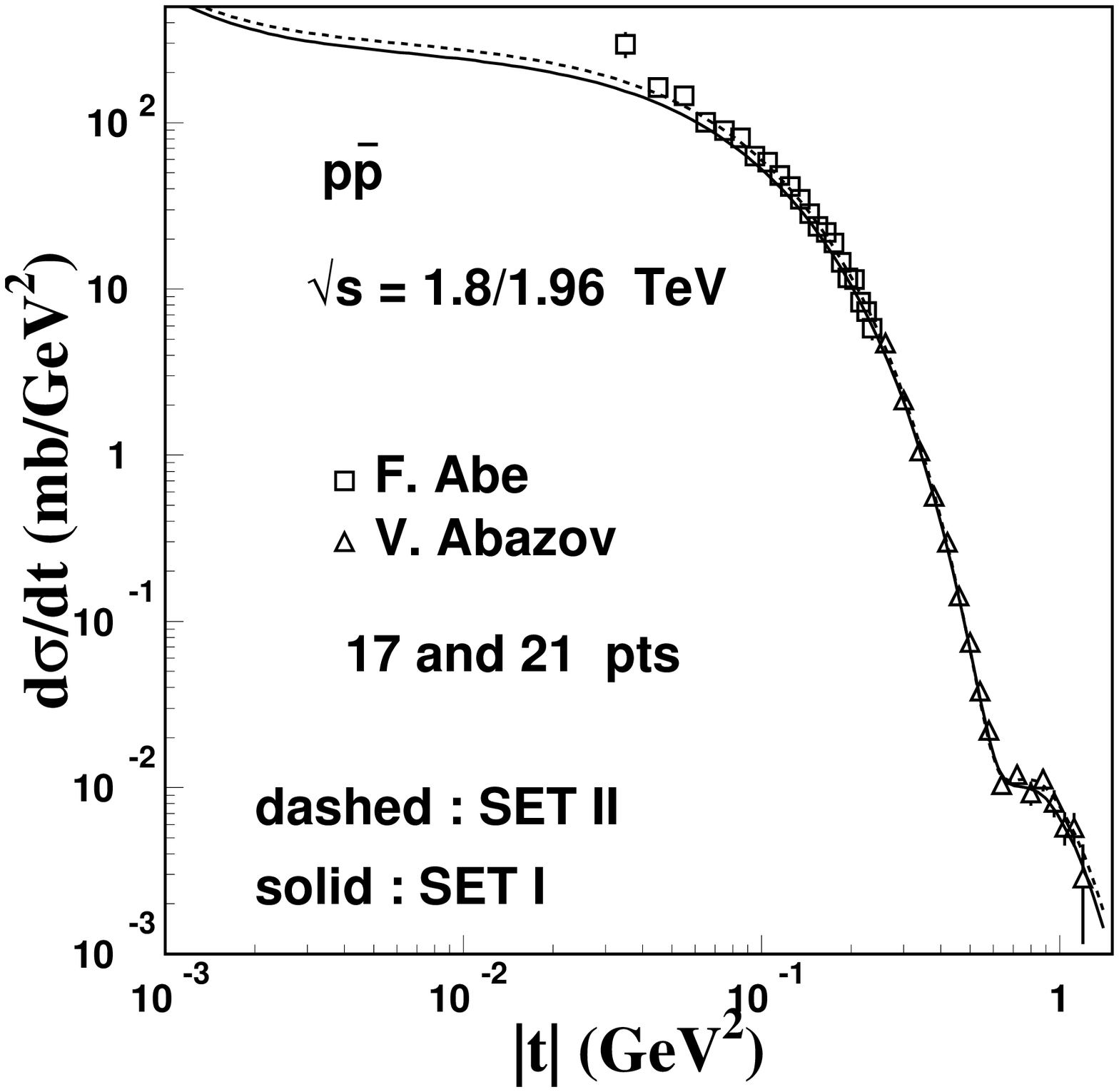}\caption{Data of SET II (21
points from CDF and 17 points from D0 experiment), with plots that enhance the
forward range. Note the smooth connection of the highest $\left\vert
t\right\vert $ CDF points with the recent D0 data. The solid and dashed lines
refer to the fitting solutions obtained with sets I and II respectively, with
parameters given in Table \ref{fittings}. In spite of the apparent proximity,
the lines lead to remarkably different values for the total cross section. The
solution for SET III falls between these two drawn lines (see dotted line in
Fig. \ref{data-figures}) and is not included here to keep clarity. }%
\end{figure}

The results of the fittings with respect to the other free parameters are
given in Table \ref{fittings}, together with some characteristic features of
the solutions. The corresponding curves representing these fittings
 of   $d\sigma/dt$ are shown in Figs. \ref{data-figures}%
,\ref{experimental-data}, \ref{hybrid-fit}.

It is important to observe 
that the discrepancy between the CDF and E-710 data shown in the RHS of
Fig. \ref{data-figures} becomes smaller as $|t|$ increases 
and both sets of data seem to connect smoothly 
to the D0 data,  as seen in \ref{hybrid-fit}.
That is, the well-known contradiction between E-710 and CDF data becomes less
serious as $|t|$ increases, and the D0 data helps to point out the connection.
Our global $|t|$ analysis helps to  describe this connection.

 \begin{table}[ptb]
\caption{ Characteristic quantities of the all-t representation for the
amplitudes. Common values for all sets: $\rho=0.141\pm0.002$ , $B_{I}%
=16.76\pm0.04 ~\nobreak\,\mbox{GeV}^{-2}$ , $B_{R}=26.24 \pm0.39~
\nobreak\,\mbox{GeV}^{-2}$ , $\alpha_{I} = 11.620\pm0.024 ~ \nobreak\,
\mbox{GeV}^{-2}$, and choice of $\beta_{R}=1.10~ \nobreak\,\mbox{GeV}^{-2}$.
The remaining free parameters are $\beta_{I}$, $\lambda_{R}$, $\sigma$. The
error bars are given by the CERN Minuit Program. SET I is built with E-710 (35
points) and D0 (17 points) data. SET II is built joining CDF (21 points) and
D0 (17 points). The complete SET III puts together CDF (26 points), E-710 (35
points) and D0 (17 points) data. $|t|_{\mathrm{infl}}$ is the position of the
inflection point in $d\sigma/dt$. $\langle\chi^{2}\rangle$ is the average
value of the squared relative theoretical/experimental deviations. }%
\label{fittings}%
\tabcolsep=0.009cm
\begin{tabular}
[c]{ccccccccc}\hline
SET~~ & ~ N & ~~$\beta_{I} $ ~~ & $\lambda_{R} $ & $|t|_{\mathrm{infl}}$ &
$~(d\sigma/dt)_{\mathrm{infl}}$ & ~ $~ \sigma$(el) & $\sigma~$ & $\langle
\chi^{2}\rangle$\\
& points & GeV$^{-2}$ & GeV$^{-2}$ & GeV$^{2}$ & mb/GeV$^{2}$ & mb & mb &
\\\hline
~I~ & ~ 52 & $~~~~ 3.7785\pm0.0078~~ $ & $~~ 3.6443\pm0.0093 ~~$ & 0.745 &
0.01013 & 16.67 & $~ 72.76\pm0.13 $ & ~ 0.7661\\
~II~ & ~ 38 & $~~~~ 3.5686\pm0.0186 ~~ $ & $~~ 3.8645 \pm0.0093 ~~$ & 0.727 &
0.01114 & 18.92 & $~ 77.63\pm0.44 $ & ~ 1.4961\\
~III~ & ~ 78 & $~~~~ 3.7441\pm0.0080 ~~ $ & $~~ 3.6784 \pm0.0096 ~~$ & 0.741 &
0.01029 & 17.02 & $~ 73.54 \pm0.20 $ & ~ 2.6591\\\hline
\end{tabular}
\end{table}

\clearpage

We recall that the above analysis is based on analytical expressions
for the scattering amplitudes applied to all $|t|$.
In the present 1.8 TeV case, the integrated use of all-$|t|$  
data is crucial since there are no data points in the very forward range,
$10^{-3}$ to $10^{-2}$ GeV$^{2}$, and the pure exponential forms are
not at all reliable. Due to the very large energy gaps in the experimental 
data, this energy region $\sqrt{s}~=1.8/1.96$  TeV is extremely important 
for the determination of the energy dependence of the total cross section, 
$\sigma(s)$  and hence for its extrapolation for ultra-high energies
treated by fundamental theorems. 

To show the importance of the use
of the full-$|t|$ amplitudes and full-$|t|$
data together, we test toy fits of the forward data of E-710 (35 points) and
CDF (21 points) experiments. The E-710 data are fitted with essentially the
same parameters as the full SET I, and this shows their nearly perfect
coherence, with the E-710 and D0 data matching very well when described by our
full-$|t|$ amplitudes. However, the separate treatment of the 21 points of the
CDF data leads to values of $\beta_{I}=3.7280 ~\GeV^{-2}$, 
$\lambda_{R}=3.3060 ~\GeV^{-2}$ and
$\sigma=79.00\pm0.57$ mb that are different from those of SET II in Table
\ref{fittings}, and this solution has a disastrous behavior for large $|t|$.
Thus, we conclude that, in our model, the use of the pure CDF points
for the determination of the very forward quantities seems not reliable, if 
it is not controlled by the D0 points of the larger $|t|$
domain. Thus in our analysis the construction of SET II is essential for the 
treatment of the CDF data.  

\clearpage

\section{Properties of Amplitudes \label{sec-amplitudes}}

It is general property of our scheme that the non-perturbative amplitudes 
fall-off rapidly after $|t|\approx1.5\nobreak\,\mbox{GeV}^{2}$, with the 
magnitude of the positive real part becoming dominant over the negative 
imaginary part for $|t|$ larger than about 2.5 GeV$^{2}$. The imaginary 
amplitude has only one zero, located near the inflection point 
of $d\sigma/dt$, while the real part has a
first zero at small $|t|$, obeying Martin's theorem \cite{Martin}, and a
second zero located after the imaginary zero. As the non-perturbative 
real part decreases, the perturbative tail remains, giving
to the differential cross section the characteristic   shape 
$1/|t|^{8}$, discussed by Donnachie and Landshoff \cite{DL_3g}. 
Such a general aspect of the scattering amplitudes have been well
verified at ISR and LHC energies \cite{ferreira1,KEK_2013}. The present
analysis at 1.8 TeV data  repeats this general behavior, 
as exhibited in Fig. \ref{1a}. 

\subsection{Role of Perturbative Tail in ${\rm p \bar p}$
scattering}

\begin{figure}[ptb]
\label{1a} \includegraphics[width=6.5cm]{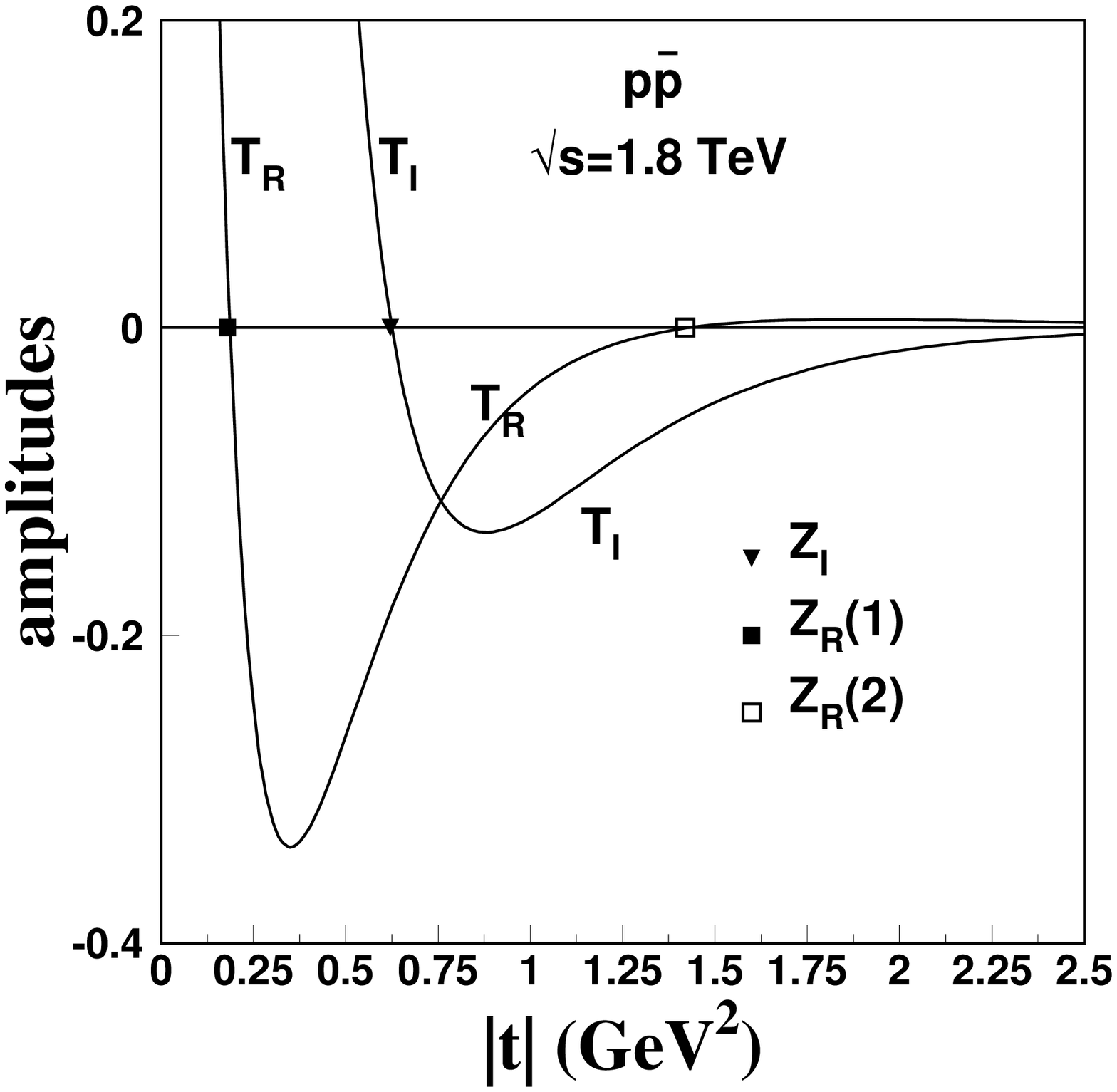}
\includegraphics[width=6.5cm]{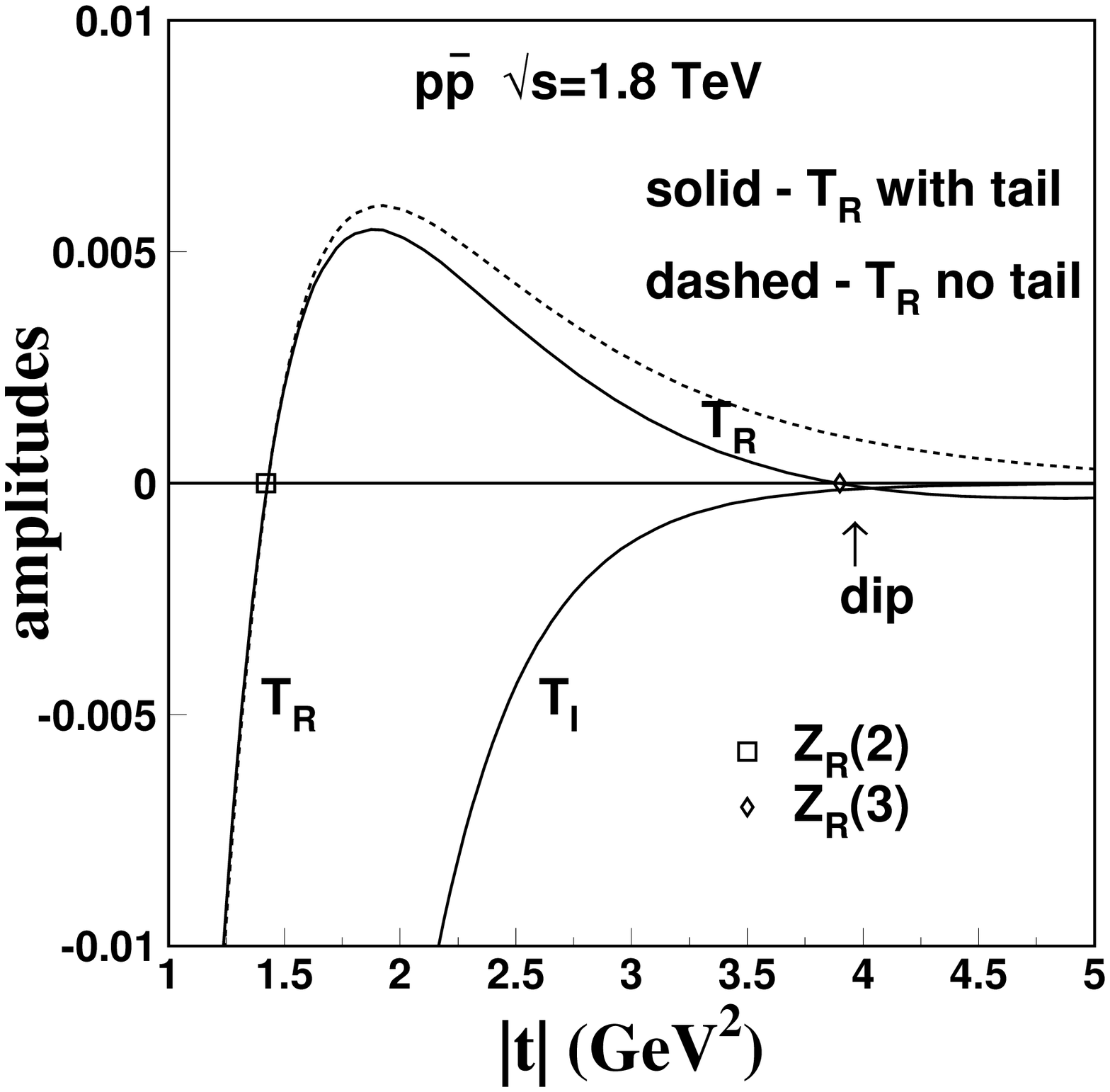}\caption{ Amplitudes in
p$\mathrm{\bar{p}}$ elastic scattering at $\sqrt{s}=1.8$ TeV shown in
different ranges and scales, described by Eqs. (\ref{real}), (\ref{imag}),
(\ref{cd1_s}) with parameters determined by phenomenology. The solid lines
drawn refer to the solutions for $T_{R}$ and $T_{I}$ obtained for SET I. In
the $|t|$ range up to about 2 GeV$^{2}$ the amplitudes are governed by
non-perturbative dynamics and are qualitativelly similar for pp and
p$\mathrm{{\bar{p}}}$, with one zero for $T_{I}$ and two zeros for $T_{R}$.
$T_{I}$ remains negative and goes fast to zero, while at $|t|\approx3$
GeV$^{2}$ the non-perturbative $T_{R}$ is positive and dominates. In
p$\mathrm{\bar{p}}$ scattering the negative contribution of the 3-gluon
exchange term inverts the sign of $T_{R}$, forming a third zero and a marked
dip in $d\sigma/dt$, with locations and depths dependent on the detail of the
$\beta_{R}$ parameter, as shown in Table \ref{dips_table}.}%
\end{figure}

  The universal (energy independent) perturbative 3-gluon
exchange process\cite{DL_3g}, given by Eq. (\ref{tail}), contributes in 
${\rm p \bar p}$  scattering with a negative sign, which 
leads to an interesting prediction. As
mentioned above, the non-perturbative real amplitude is positive in the
transition region, and the inclusion of the negative tail amplitude
leads eventually to its cancellation and the creation of a third real zero
(see Table \ref{dips_table}).
This mechanism is shown in the RHS of Fig. \ref{1a}, where we draw  
two curves for the real amplitude, with solid line and dashed line, 
corresponding respectively  to presence and absence of 
perturbative contribution. 
 
\begin{figure}[b]
\label{large_dsdt} \includegraphics[width=6.5cm]{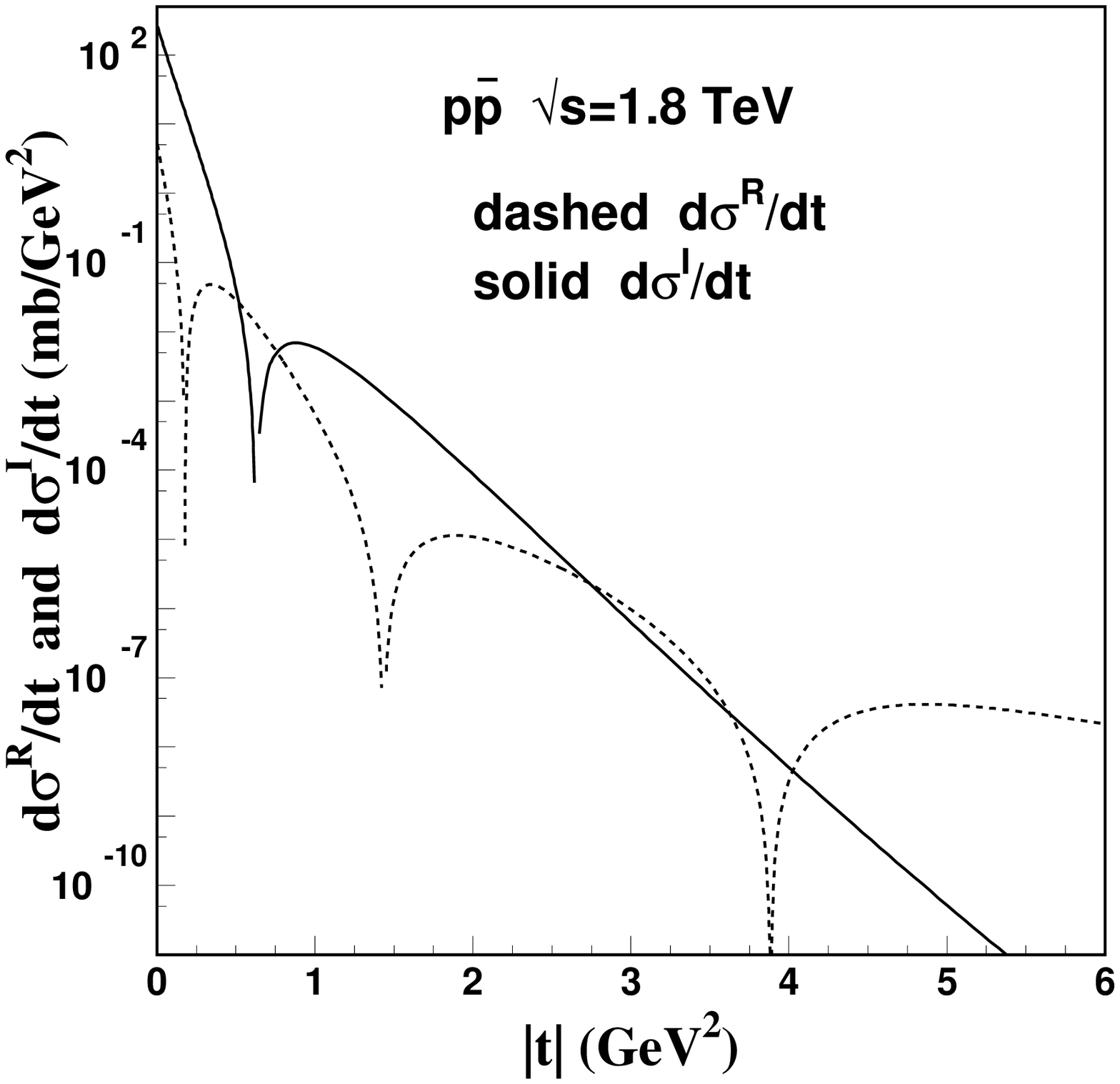}
\includegraphics[width=6.5cm]{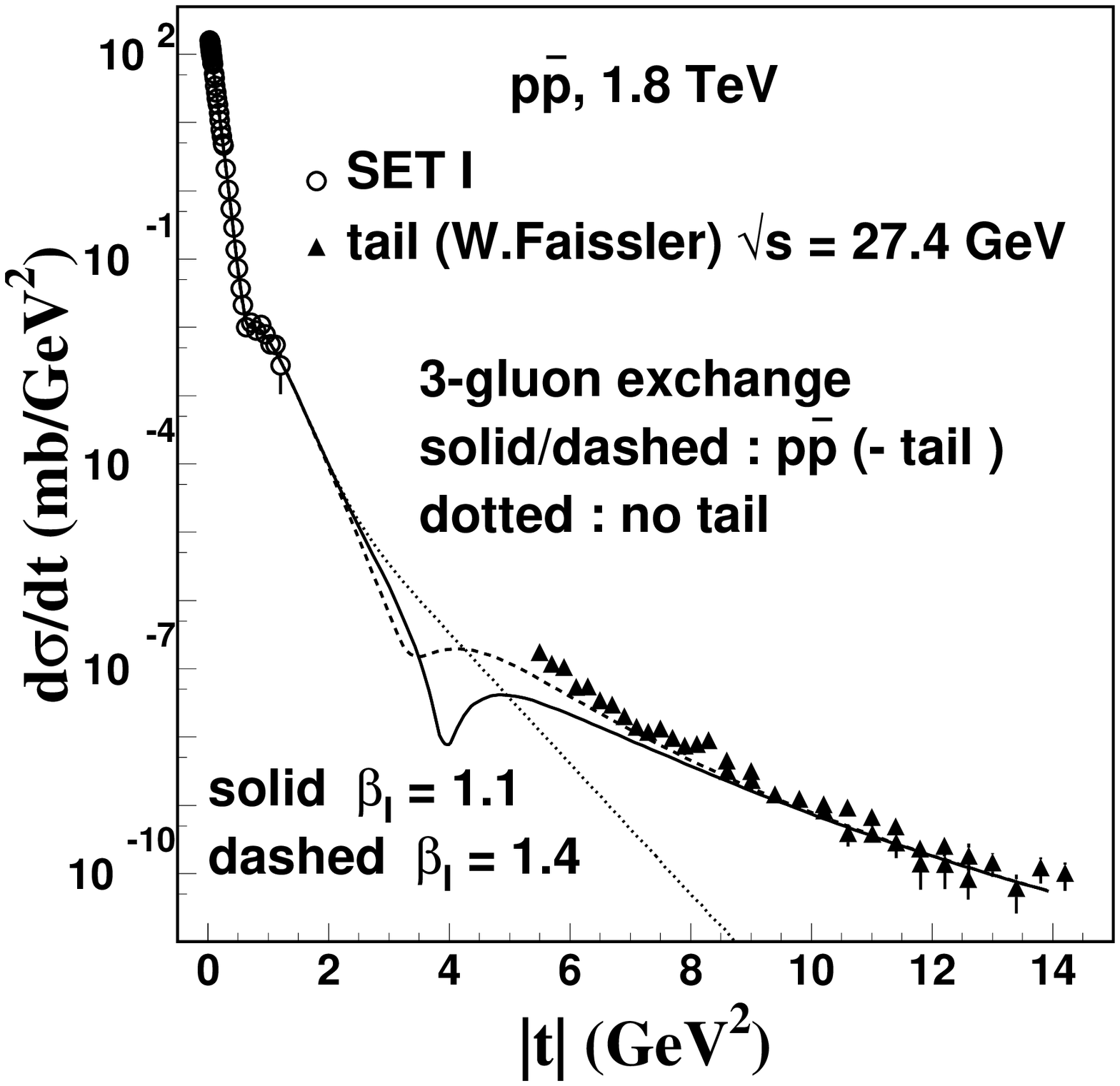}\caption{ The plots show the
predictions for the contributions of real $d\sigma^{R}/dt$ and imaginary
$d\sigma^{I}/dt$ parts of $d\sigma/dt$ in the presence of the real
perturbative tail due to 3 gluon exchange. In p$\mathrm{\bar p}$ scattering
the negative sign of the tail causes a zero in $d\sigma^{R}/dt$ and a dip in
$d\sigma/dt$ located in the range 3-5 GeV$^{2}$. The RHS figure shows two
examples of the dip structure, formed with $\beta_{R}=1.10 ~
\nobreak\,\mbox{GeV}^{-2}$ (solid) and $\beta_{R}%
=1.40~\nobreak\,\mbox{GeV}^{-2}$ (dashed) as given in Table \ref{dips_table}.
We suggest that the analysis of data from the Fermilab experiment at 1.96 TeV
be extended to investigate this dip region. }%
\end{figure}

As the imaginary part is not dominant in this region, a marked dip may 
be observed in $d\sigma/dt$. This is shown in Fig.
\ref{large_dsdt}. In this figure (RHS), we also show in dotted line 
the behavior of cross section with  non-pertubative amplitudes only,  
without the effect of pertubative tail.

The precise form of this dip-bump structure created by the
perturbative tail depends sensitively on the values of model parameters 
(such as $\beta_R$) that  
govern the properties of the transition domain. Unfortunately, 
the existing data stops at about
$|t|=1.2$ GeV$^{2}$, leaving the higher $|t|$ region without information to
fix the connection with the range of the perturbative tail. Thus the parameter
$\beta_{R}$ cannot be fixed accurately, and as its value is crucial for the
prediction of the position and depth of a dip in the transition region for
${\rm p \bar p} $  scattering at 1.8 TeV, we present in Table \ref{dips_table} 
two alternative choices, with
$\beta_{R}$ = 1.10 and 1.40 GeV$^{-2}$.

In Table \ref{dips_table} are given the values of $|t|$ at the zeros of the
amplitudes, and the locations of the dip and bump in $d\sigma/dt$ at large
$|t|$ that are due to the contribution of the three-gluon exchange term. The
quantity $\mathrm{{ratio}=(d\sigma/dt)_{{bump}}/(d\sigma/dt)_{{dip}}}$ that
informs about the shape of the structure depends strongly on the values of the
parameter $\beta_{R}$, that must be determined by experiment, 
necessarily with extension of the measured range to higher $|t|$ values. 
The common parameters are given
in Eq. (\ref{basic_parameters}).  The fitting of each solution is needed only to
evaluate $\lambda_{R}$.

\begin{table}[ptb]
\caption{ Positions of zeros of the real and imaginary amplitudes, locations
of the dip and bump at large $|t|$ predicted by the introduction of the
perturbative tail of negative sign, and the ratio characterizing the shape of
this structure. The parameter $\beta_{R}$, that determines the behavior of the
real part at the end of the non-perturbative region, is not tightly determined
by the data (that ends at 1.2 GeV$^{2}$), and has important role for the
location and depth of the large $|t|$ dip. We present results for two choices
of $\beta_{R}$. The parameter $\lambda_{R}$ varies in the fits, following the
choice of $\beta_{R}$. The quantities $\rho$, $B_{I}$ , $B_{R}$ , $\alpha_{I}$
are universal, as in Table \ref{fittings}. The quantity ratio is
$[d\sigma/dt]_{\mathrm{bump}}/[d\sigma/dt]_{\mathrm{dip}}$. }%
\label{dips_table}
\begin{tabular}
[c]{cccccccccc}\hline
SET & $\beta_{R}$~ & $\lambda_{R}$ & $\mathrm{Z_{I}} $ ~ & $\mathrm{Z_{R}(1)}%
$~ & $\mathrm{Z_{R}(2) }$~ & $\mathrm{Z_{R}(3)}$~ & $|t|_{\mathrm{dip}}$ &
$|t|_{\mathrm{bump}}$ & ratio\\
& GeV$^{-2}$ & GeV$^{-2} $ & GeV$^{2}$ & GeV$^{2}$ & GeV$^{2}$ & GeV$^{2}$ &
GeV$^{2}$ & GeV$^{2}$ & \\\hline
I & 1.10 & 3.6443 & 0.6253 & 0.1771 & 1.4336 & 3.8827 & 3.9456 & 4.8631 &
5.4567\\\hline
I & 1.40 & 3.6328 & 0.6253 & 0.1776 & 1.5884 & 3.0605 & 3.4839 & 4.1212 &
1.3086\\\hline
II & 1.10 & 3.8645 & 0.6156 & 0.1792 & 1.2986 & 4.3159 & 4.3520 & 5.3314 &
8.4118\\\hline
II & 1.40 & 3.8492 & 0.6156 & 0.1799 & 1.4217 & 3.3047 & 3.6434 & 4.2920 &
1.3761\\\hline
III & 1.10 & 3.6784 & 0.6231 & 0.1776 & 1.3987 & 3.9781 & 4.0312 & 4.9580 &
6.2212\\\hline
III & 1.40 & 3.6662 & 0.6231 & 0.1781 & 1.5452 & 3.1181 & 3.5111 & 4.1609 &
1.3442\\\hline
\end{tabular}
\end{table}

\clearpage

\subsection{Comparison with the BSW model \label{BSW-model}}

In our work we emphasize the importance of determination of the
amplitudes that describe the observed quantities in elastic scattering.
However, this description is naturally model dependent, so that it is interesting
to compare our results to those of other models. Results on amplitudes that
can be directly compared with ours are given by the model proposed by
Bourrely, Soffer and Wu (hereafter referred to as BSW model) \cite{BSW}. The
comparison is presented below.

The parameters of BSW model at 1.8 TeV are :
\[
\sigma=73.99 ~\mathrm{mb} ~ ; ~ \rho=0.129 ~; ~ B_{I}=18.12 ~
\nobreak\,\mbox{GeV}^{-2} ~ ; ~ B_{R}=22.82 ~ \nobreak\,\mbox{GeV}^{-2} ;
\]
\[
Z_{I} = 0.685 ~ \nobreak\,\mbox{GeV}^{2} ~ ; ~ Z_{R}(1) = 0.275 ~
\nobreak\,\mbox{GeV}^{2} ~;~ Z_{R}(2) = 2.040 ~ \nobreak\,\mbox{GeV}^{2} ~ .
\]

Instead of the inflection points, the model gives the first dip and bump 
 for $d\sigma/dt$,  with a flat structure,  with the values 
\[
|t|_{\mathrm{dip}} = 0.72 ~ \nobreak\,\mbox{GeV}^{2} ; ~ |t|_{\mathrm{bump}} =
0.90 ~ \nobreak\,\mbox{GeV}^{2} ~ ;~ \mathrm{ratio} = 1.226~.
\]

\begin{figure}[ptb]
\caption{Cross-section - comparison with BSW. In the range 1 - 2 GeV$^{2}$,
the BSW model has both real and imaginary amplitudes with magnitudes larger
than ours (see Fig. \ref{BSW-amplitudes} , with consequence that the 
dotted line is higher.
}%
\label{BSW-dsigdt}%
\includegraphics[width=7.5cm]{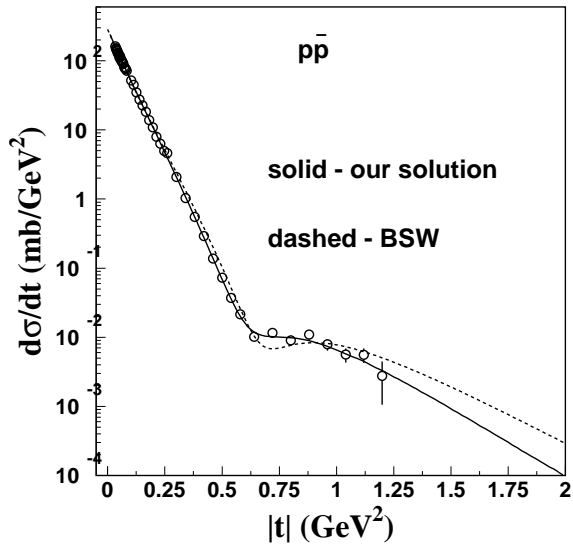}
\end{figure}

\begin{figure}[ptb]
\caption{Comparison of BSW amplitudes with ours. As shown in the LHS plot, the
amplitudes of BSW are qualitatively similar to ours. In the RHS plot  
the  vertical scale is amplified to illustrate the difference in the large
$|t|$ domain.   }%
\includegraphics[width=6.5cm]{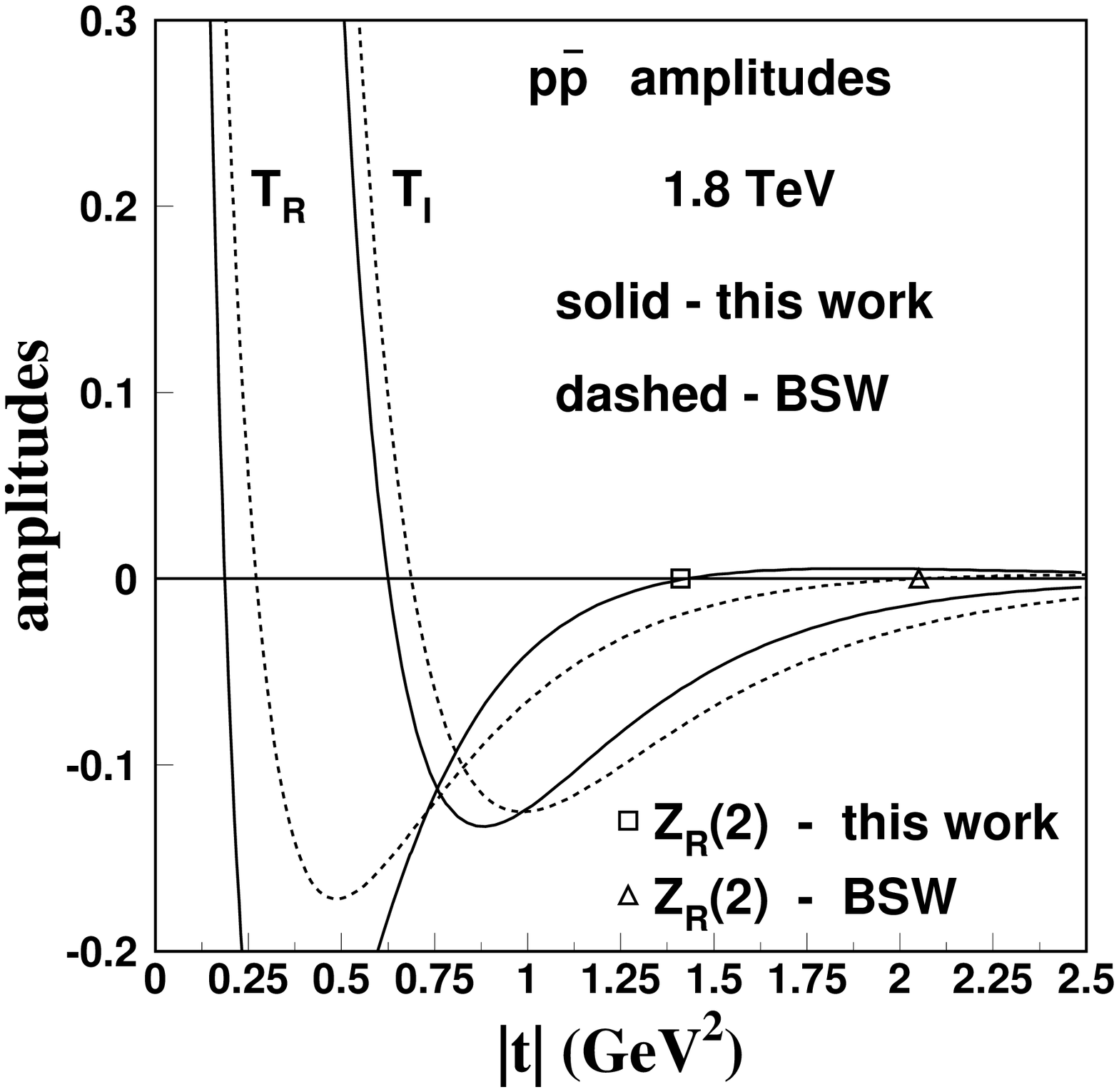}
\includegraphics[width=6.5cm]{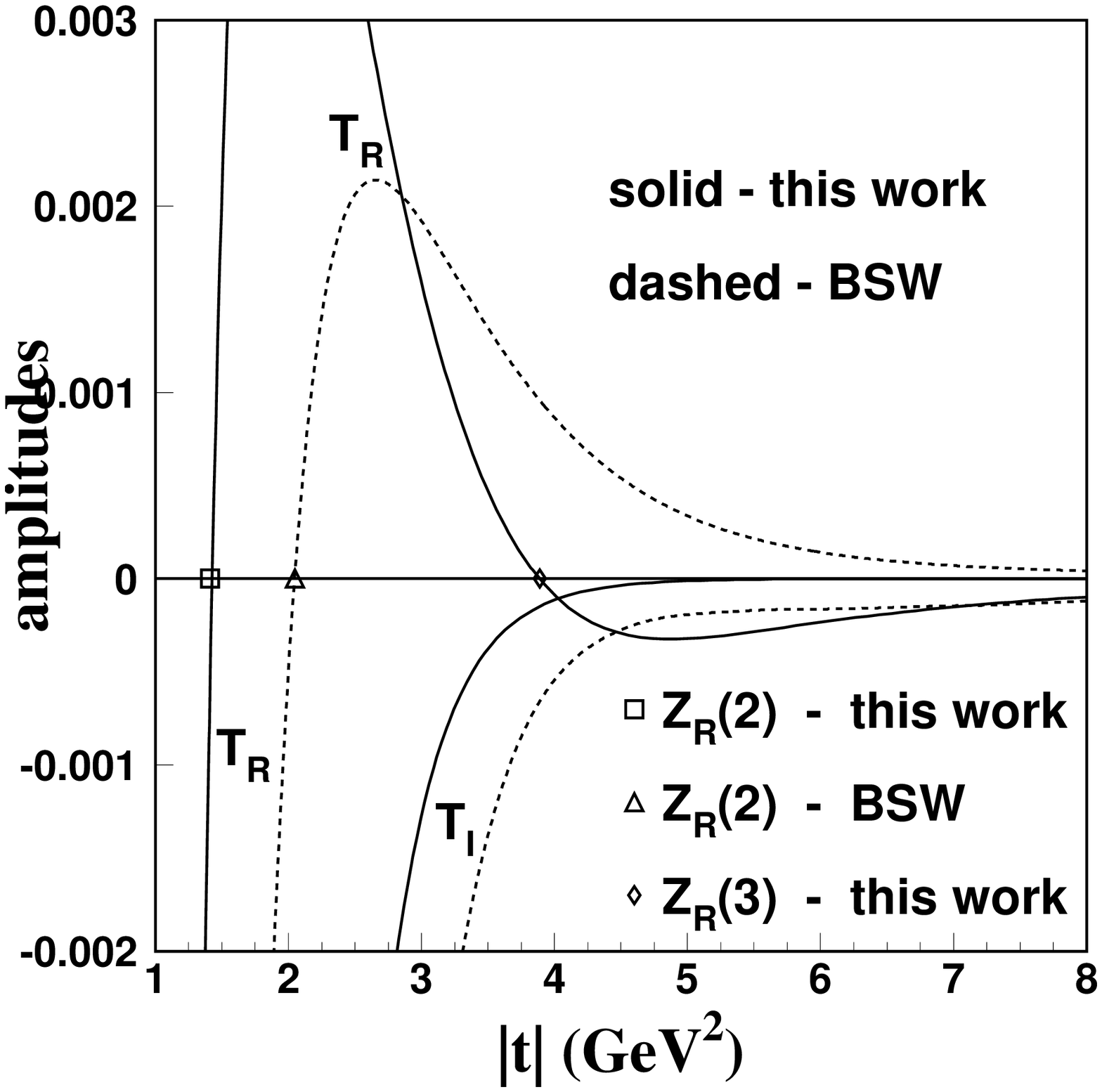}
\label{BSW-amplitudes}%
\end{figure}

To compare the BSW model with our calculations, we show in Fig.
\ref{BSW-dsigdt} the comparison of cross-sections, and in Fig.
\ref{BSW-amplitudes} the comparison of amplitudes.

As shown in the LHS plot, the amplitudes of the BSW model are
qualitatively similar to ours in the low and mid $|t|$ ranges, with one
imaginary and two real zeros, all of which occur at higher $|t|$ than ours. At
higher $|t|,$ shown in the RHS plot, the important difference appear. In BSW
the imaginary magnitude dominates and falls to zero more slowly. In our case
the real part determines the tail behavior. At very large $|t|$ (namely
$|t|\geq6~\nobreak\,\mbox{GeV}^{2}$) the roles of the imaginary and real
magnitudes are interchanged in comparison to ours. These qualitative 
similarities and differences of
the two models have also been observed also in 7 TeV case \cite{KEK_2013}.

\clearpage

\section{Summary and Discussion  {\label{sec-conclusions} }  }

In this work we present precise descriptions of the elastic
scattering amplitudes and of the differential cross sections  
for the p\={p} collisions, merging the recent 1.96 TeV
and the former 1.8 TeV data. We use analytical forms for the 
real and imaginary amplitudes covering the full $|t|$ range, identifying  
their zeros, signs, ranges of dominance and the
interplays that fix the observed details. 
To investigate the existing discrepancy of  the 1.8 TeV data of the 
E-710 and CDF experiments in the presence of the new 1.96 data, 
we construct three different combinations of data for the 
evaluation of total cross section and for the representation of 
the differential cross sections. Results are given in Tables 
\ref{fittings} and \ref{dips_table}, and the solutions are illustrated 
in Figs. \ref{data-figures}, \ref{experimental-data}, \ref{hybrid-fit}
for the data and in Fig. \ref{1a} for the amplitudes.

Fig.\ref{hybrid-fit} clearly shows how delicate is the extrapolation of
experimental data of $d\sigma/dt$ towards $|t|=0$. 
Particularly in the case where data points are lacking in the forward 
region,  a more structured approach is fundamental 
for the determination of the so called forward
scattering parameters $\sigma$, $\rho$, $B_{I}$, $B_{R}$ 
+,
since they can only be defined  in the limit $|t|\rightarrow 0$. 
Obviously we cannot avoid model-dependence,  
but we believe that the general features of the real and imaginary 
amplitudes such as magnitudes, curvatures, zeros and signs are fundamental 
and should be incorporated in the analysis of the data. 
For example, the usually adopted  assumption $B_{R}=B_{I}$ 
is essentially wrong and may lead to incorrect values for the 
forward scattering parameters.

Our work revises the values of total cross section and slope parameters 
that are reported in the literature, suggesting new values, which we 
believe to be more realistic. 
 In addition, we show that with the use of the hybrid set combining 
CDF with large-$|t|$ D0 data, the well-known discrepancy of CDF and 
E-710 data can be more tamed. 

We show that, as is the cases of SPS and LHC energies\cite{ferreira1,
KEK_2013}, the universality of the perturbative three-gluon exchange 
tail as asymptotic behavior of the real part is consistent 
with the data, and in the particular ${\rm p \bar p}$ case, leads to a very
interesting consequence, due to the sign of this contribution. For
$|t|>4 ~\GeV^2$,  the non-perturbative real amplitude is positive and 
dominates the negative imaginary amplitude. The inclusion of the negative 
real amplitude of the pertubative tail makes the
real amplitude eventually negative again, creating a third zero.
As the imaginary part is not dominant there, a marked dip may appear in
$d\sigma/dt$  in this transition region as shown in Fig.
\ref{large_dsdt}. As mentioned in the text, the precise form of this dip
structure depends on the parameters which govern the behavior 
data in the transition region between non-perturbative and perturbative 
dominances.  

The confirmation of the presence  of  this dip 
in the  $3\lesssim |t| \lesssim4 ~ \GeV^2$ range
would characterize
the sign of the real amplitude and its dominance over the imaginary part 
in the mid-$t$ region, thus giving  model-free information on 
the elastic scattering amplitude.
We then  propose the analysis  of the  
collected data of the D0 
collaboration at values of $|t|$ beyond those already published.

\section{Acknowledgments}

The authors wish to thank CNPq, PRONEX and Faperj for financial support. A
part of this work has been done while TK stayed as a visiting professor at
EMMI and FIAS at Frankfurt. TK expresses his thanks to the hospitality of
Profs. H. Stoecker and D. Rischke. Conversations with M. Rangel and G. Alves,
of the D0 Collaboration, are gratefully acknowledged.




\end{document}